\documentclass[twocolumn,showpacs,aps,prd,superscriptaddress]{revtex4}

\usepackage{graphicx}
\usepackage{dcolumn}
\usepackage{amsmath}
\usepackage{epsfig}

\input pubboard/babarsym

\def\pstar     {\ensuremath {p^*}\xspace} 
\def\ldata     {\ensuremath {124 \invfb}}

\def\etal {{\it et al.}}

\def\Thc      {\ensuremath {\Theta_c(3100)^0}\xspace} 
 
\def\dstm     {\ensuremath {D^{*-}}\xspace} 
\def\pdstm    {\ensuremath {p D^{*-}}\xspace} 
\def\Thetaplus{\ensuremath {\Theta(1540)^+}\xspace} 
\def\Ximm     {\ensuremath {\Xi(1860)^{--}}\xspace} 
\def\Xizero   {\ensuremath {\Xi(1860)^0}\xspace}

\def\eeqq     {\ensuremath {e^+e^-\! \to q\overline{q}}\xspace}
\def\eecc     {\ensuremath {e^+e^-\! \to c\overline{c}}\xspace}
\def\dbar0    {\ensuremath {\overline{D}^0}\xspace}
\def\kpi      {\ensuremath {K^+\pim}\xspace}
\def\ktpi     {\ensuremath {K^+\pim\pip\pim}\xspace}
\def\kpis     {\ensuremath {K^+\pim\pi^-_s}\xspace}
\def\ktpis    {\ensuremath {K^+\pim\pip\pim\pi^-_s}\xspace}
\def\pis      {\ensuremath {\pi^-_s}\xspace}
\def\mkpi     {\ensuremath {m_{K^+\pim}}\xspace}
\def\mkpis    {\ensuremath {m_{K^+\pim\pis}}\xspace}
\def\mktpi    {\ensuremath {m_{K^+\pim\pip\pim}}\xspace}
\def\mktpis   {\ensuremath {m_{K^+\pim\pip\pim\pis}}\xspace}

\def\mkttpis  {\ensuremath {m_{K^+\pim(\pip\pim)\pis}}\xspace}
\def\mpktpis  {\ensuremath {m_{pK^+\pim(\pip\pim)\pis}}\xspace}
\def\mpdst    {\ensuremath {m_{pD^*}}\xspace}
\def\hwid     {28~\mev} 
\def\sdotB    {\ensuremath {{\cal B}\!\cdot\! \sigma }\xspace}
\def\dsdpst   {\ensuremath {{\rm d}\sigma /{\rm d}\pstar}\xspace}
\def\dsdotB   {\ensuremath {{\cal B}\!\cdot\! {\rm d}\sigma /{\rm d}\pstar}\xspace}

\newcommand{\BABARPubYear}    {06}
\newcommand{\BABARPubNumber}  {007}

\newcommand{\SLACPubNumber} {11779}
\newcommand{\LANLNumber} {0604006}

\def\figurebox#1#2#3{%
    \def\arg{#3}%
    \ifx\arg\empty
    {\hfill\vbox{\hsize#2\hrule\hbox to #2{\vrule\hfill\vbox to #1{\hsize#2\vfill}\vrule}\hrule}\hfill}%
    \else
    {\hfill\epsfbox{#3}\hfill}%
    \fi}

\begin{document}

\preprint{\babar-PUB-\BABARPubYear/\BABARPubNumber} 
\preprint{SLAC-PUB-\SLACPubNumber} 

\begin{flushleft}
SLAC-PUB-\SLACPubNumber\\
\babar-PUB-\BABARPubYear/\BABARPubNumber\\
hep-ex/\LANLNumber\\
\end{flushleft}

\title{
{\large \bf
Search for the Charmed Pentaquark Candidate
\boldmath{$\Theta_c(3100)^0$} in \boldmath{$e^+e^-$} Annihilations at 
\boldmath{$\sqrt{s}=10.58$} \gev 
} }

%
\author{B.~Aubert}
\author{R.~Barate}
\author{M.~Bona}
\author{D.~Boutigny}
\author{F.~Couderc}
\author{Y.~Karyotakis}
\author{J.~P.~Lees}
\author{V.~Poireau}
\author{V.~Tisserand}
\author{A.~Zghiche}
\affiliation{Laboratoire de Physique des Particules, F-74941 Annecy-le-Vieux, France }
% ...
\collaboration{The \babar\ Collaboration}
\noaffiliation

\date{\today}

\begin{abstract}
\noindent
We search for the charmed pentaquark candidate reported by the H1
collaboration, the \Thc, in
$e^+e^-$ interactions at a center-of-mass (c.m.) energy of 10.58 GeV, using
\ldata of data recorded with the \babar\ detector at
the PEP-II $e^+e^-$ facility at SLAC.
We find no evidence for such a state in the same \pdstm decay
mode reported by H1,
and we set limits on its production cross section times branching
fraction into \pdstm as a function of c.m.\ momentum.
The corresponding limit on its total rate per \eeqq event, times branching
fraction, is about three orders of magnitude lower than rates
measured for the charmed $\Lambda_c$ and $\Sigma_c$ baryons in such events.

\end{abstract}

\pacs{13.25.Hw, 12.15.Hh, 11.30.Er}

\maketitle

\noindent
Ten experimental groups have recently reported narrow 
enhancements near 1540~\mevcc
in the invariant mass spectra for $n K^+$ or 
$p \KS$~\cite{thetarefs}.
The minimal quark content of a state that decays 
strongly to $n K^+$ is $dduu\overline{s}$;
therefore, these mass peaks have been interpreted as a 
possible pentaquark state, called \Thetaplus.
The NA49 experiment has reported narrow 
enhancements near 1862~\mevcc in the 
invariant mass spectra for $\Xi^- \pi^-$ and $\Xi^- \pi^+$~\cite{prl92:042003};
the former has minimal quark content $dssd\overline u$,
and these two mass peaks have also been interpreted as  
possible pentaquark states, named \Ximm and \Xizero 
[also known as $\Phi(1860)$], 
with the latter 
being a mixture of $ussu\overline u$ and $ussd\overline d$.
The H1 experiment has reported a narrow enhancement at a mass of 
3099$\pm$6~\mevcc in the mass spectrum for \pdstm~\cite{h1plb},
which has a minimal quark content of $uudd\overline c$, making this 
a possible charmed pentaquark state, named \Thc.
On the other hand, there are numerous experimental searches with
negative results~\cite{review}:
several experiments observe large samples of strange baryons with mass 
similar to that of the \Thetaplus, e.g. $\Lambda(1520) \!\!\to\! p K^-$,
but no evidence for the \Thetaplus;
several observe large samples of the nonexotic $\Xi^-$ baryon, but not 
the \Ximm or \Xizero states;
and several with large samples of \dstm do not observe the \Thc state.
Our recent search~\cite{bbrlimits} for the \Thetaplus and \Ximm in $e^+e^-$
annihilations found no evidence for these states, and we set limits on their
production rates in \eeqq events of factors of eight and four, respectively,
below rates expected for ordinary baryons of the same masses.

Here we report the results of an inclusive search for the charmed
pentaquark candidate \Thc in $e^+e^-$ annihilation data;
we expect equal production of the charge conjugate state, and its 
inclusion is implied throughout this article.
The data were recorded with the \babar\ detector~\cite{BABARdetector} 
at the PEP-II asymmetric-energy $e^+e^-$ storage rings 
located at the Stanford Linear Accelerator Center.
The data sample represents an integrated luminosity of \ldata\
collected at an $e^+e^-$ c.m.\ energy at or just below 
the mass of the \Y4S resonance.
We study the same decay mode as in the H1 analysis, $\Thc \to \pdstm$, 
where the \dstm decays to $\dbar0 \pis$
(\pis denotes a ``slow" pion from the \dstm decay),
and the \dbar0 decays to \kpi.
In addition, we consider the mode in which  the \dbar0 decays to \ktpi.

The \babar\ detector is described in detail in Ref.~\onlinecite{BABARdetector}.
We use all events accepted by our trigger,
which is  more than 99\% efficient for both \eeqq
and $e^+e^-\! \to \Y4S$ events.
We use charged tracks reconstructed in the five-layer silicon vertex tracker
(SVT) and the 40-layer drift chamber (DCH).
The combined momentum resolution, $\sigma(p_T)$, is
given by
$[\sigma(p_T) / p_T]^2 = [0.0013 p_T]^2 + 0.0045^2$,
where $p_T$ is the momentum transverse to the beam axis measured in \gevc.
Particles are identified as pions, kaons, or protons with a combination of 
the energy-loss measured in the two tracking detectors and 
the Cherenkov angles measured in the detector of internally reflected 
Cherenkov radiation (DIRC).

We evaluate the \Thc reconstruction efficiency and invariant mass resolution 
from two simulations.
For production in \eecc events, we use the JETSET~\cite{jetset} Monte Carlo
generator with the mass and width of the $\Sigma_c(2455)^0$ baryon set
to 3099~\mevcc and 1~\mev, respectively, and allow only the \pdstm decay mode.
We leave all other parameters unchanged, and a momentum spectrum 
similar to those of nonexotic charmed baryons is produced.
The events have a total charm of $\pm$2, but this has negligible
effect on the number and distribution of additional particles in the event,
which are the quantities of interest here.
We also simulate \Y4S decays in which one $B$ decays generically in our
standard framework~\cite{evtgen} and the other decays 
into a state containing a $\Sigma_c(2455)^0$ with
parameters adjusted in the same way.
This gives a much softer momentum spectrum, cut off at the
kinematic limit for $B$ meson decays, and a different
environment in terms of other particles in the event.
We find that the efficiency and resolution depend primarily on 
the \Thc momentum and polar angle in the laboratory frame, and negligibly on
other aspects of the production process or event environment.
We use large control samples of particles identified 
in the data to correct small 
inaccuracies in the performance predicted by the 
GEANT-based~\cite{GEANT} detector simulation.

We choose \Thc candidate selection criteria designed for high efficiency and
low bias against any production mechanism.
We use charged tracks reconstructed with at least twelve coordinates
measured in the DCH, and select identified pions, kaons and protons.
The identification criteria for pions and kaons are fairly loose,
having efficiencies better than 99\% and misidentification rates
below 1\% for momenta below 0.5~\gevc where energy loss in the SVT and
DCH provide good separation, 
and efficiencies of roughly 80\% and misidentification rates
below 10\% for momenta above 0.8~\gevc where the Cherenkov angles are
measured well in the DIRC.
The criteria for identified protons are tighter.
For momenta below 1~\gevc and above 1.5~\gevc
the efficiencies are better than 95\% and 75\%, and the
misidentification rates are below 1\% and 3\%, respectively.

In each event we consider every combination of identified $p K^+ \pim\pim$ and 
$p K^+ \pim\pip\pim\pim$ and perform a topological fit to each
combination with the hypothesized decay chain
$X\!\!\to \!\pdstm \!\!\to\! p\dbar0 \pis \!\!\to\! pK^+ \pim(\pip\pim)\pis$.
No mass constraints are used in the fit, but the decay products at
each stage are required to originate at a single space point.
The \dbar0 has a finite flight distance, and we require the confidence
level of the $\chi^2$ for its decay vertex to exceed $10^{-4}$.

\begin{figure}[t]
\begin{flushright}
\hspace*{-0.3cm}
\includegraphics[width=4.3cm]{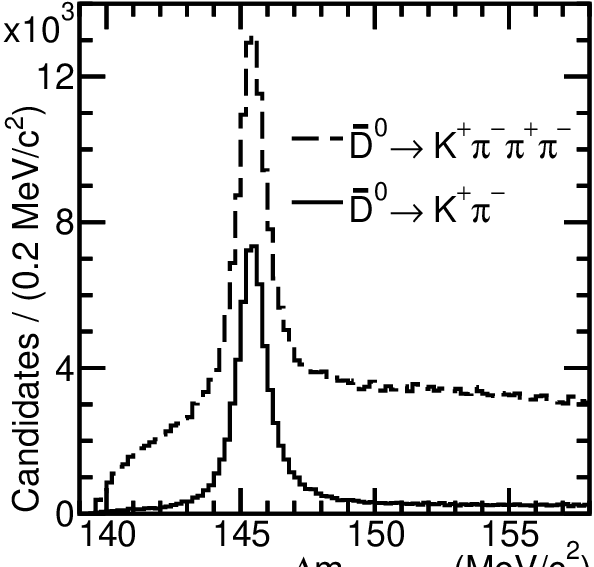}
\includegraphics[width=4.3cm]{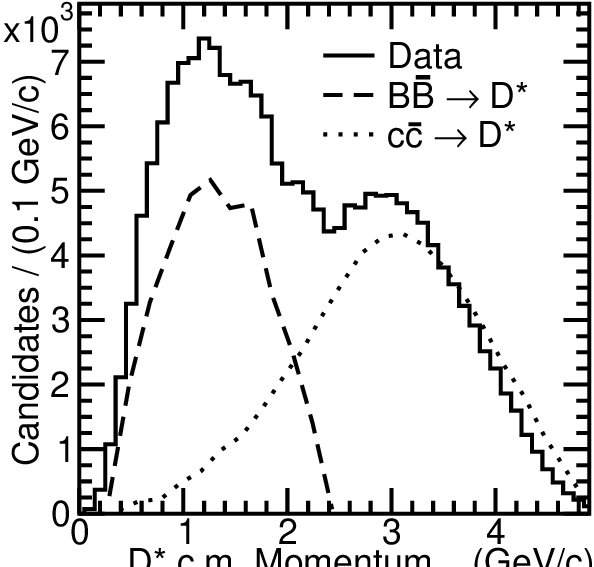}
\end{flushright}
\vspace{-0.6truecm}
 \caption{
  a)
  Distributions of the invariant mass difference 
  between the \dstm and \dbar0 combinations in
  $X \!\!\to\! \pdstm$ candidates in the data, for the two \dbar0
  decay modes.
  b)
  Distribution of the reconstructed momentum of the \dstm in the
  \epem c.m.\ frame for \pdstm candidates in the data (histogram);
  the dashed (dotted) line represents the \dstm spectra measured in 
  \Y4S decays (\eecc events) scaled as described in the text.
 }
 \label{fig:mdiffs}
\end{figure}

We select candidates in which both the reconstructed \dbar0 and \dstm
masses are within $20 \,\mevcc$ of the peak value,
namely 1843.8$\,<\! m_{K^+\pim(\pip\pim)}\!\! <\,$1883.8~\mevcc and 
1989$\,<\! m_{K^+\pim(\pip\pim)\pi^-_s}\!\! <\,$2029~\mevcc.
In Fig.~\ref{fig:mdiffs}a we show the distributions of the
differences in reconstructed invariant mass 
$\Delta m\!\! =\!\! \mkpis \! - \mkpi$ and
$\mktpis\! - \mktpi$ for these
$X \!\to\! p\kpis$ and 
$p\ktpis$ candidates, respectively.
Clear signals for \dstm are visible in both cases, with peak
positions and widths ($\sim$0.6~\mevcc) 
consistent with expectations from our simulation.
The widths ($\sim$6~\mevcc) of the corresponding \dbar0 and
\dstm peaks (not shown) are underestimated by about 10\% in the simulation.
We require a mass difference within 2~\mevcc of the 
peak value, 143.48$<\! \Delta m\! <$147.48~\mevcc.

About 55,000 $D^{*-} \!\!\to\! K^+\pim\pis$ decays and
73,000 $D^{*-} \!\!\to\! K^+\pim\pim\pip\pis$ decays
are present in the selected data over respective backgrounds of 4,000
and 62,000 random combinations.
No event in either the data or simulation has more than one surviving 
\pdstm candidate.
Without the proton requirement, over 750,000 \dstm are seen.
Figure~\ref{fig:mdiffs}b shows the distribution of the \dstm momentum, 
\pstar, in the c.m.\ frame for the selected data.
A characteristic two-peak structure is evident, in which the peak at
lower \pstar values is due to \dstm from decays of $B$ hadrons from
\Y4S decays,
and the peak at higher \pstar values is due to \eecc events.
For purposes of illustration, we show the spectra 
measured~\cite{bellelc} from these two sources on Fig.~\ref{fig:mdiffs}b,
scaled by our integrated luminosity, average efficiency and 
fraction of events with a proton.
The shape is modified by the selection criteria; 
in particular, the proton
requirement shifts the edge at the highest \pstar values.
The background is verified by sideband studies to be concentrated at 
lower \pstar values;
it is clear that we are sensitive to \Thc production from both of
these sources.

We evaluate the \Thc reconstruction efficiency for each
search mode from the simulation, as a function of \pstar.
High-mass particles at low \pstar are boosted forward in our laboratory
frame, so that the probability of losing at least one track outside
the acceptance is large, and the efficiencies are low, about
10\% and 5\% for the \kpi and \ktpi modes, respectively.
The efficiencies rise with increasing \pstar to respective maximum values of 
30\% and 22\% at the kinematic limit.
The invariant mass requirements introduce negligible signal loss.
The relative systematic uncertainties on the 
tracking and particle identification efficiencies total 6--8\%;
at low and high \pstar values, there is a contribution of similar size
from the statistics of the simulation.

We calculate the \Thc candidate invariant mass as
$\mpdst \!\!\equiv\! \mpktpis \!\! -\!\mkttpis\!\! +\! m_{\dstm}$, where
$m_{\dstm}\!\! =\,$2010~\mevcc is the known \dstm mass~\cite{RPP2004}.
We take the resolution on this quantity from the simulation, as it is
insensitive to the simulated $D^{(*)}$ mass resolution and
previous studies involving protons combined with \KS~\cite{bbrlimits}
showed the proton contribution to be well simulated.
We describe the resolution by a sum of two 
Gaussian functions with a common center.
The width of the core (tail) Gaussian averages 2.5~(20)~\mevcc, almost
independent of \pstar, and the wider Gaussian contributes between 20\%
of the total at low \pstar and 10\% at high \pstar.
The overall resolution, defined as the FWHM of the resolution function
divided by 2.355, averages 2.8 and 3.0~\mevcc for the \kpi and \ktpi decay
modes, respectively, with a small dependence on \pstar.

\begin{figure}[t]
\begin{flushright}
\hspace*{-0.3cm}
\includegraphics[width=8.7cm]{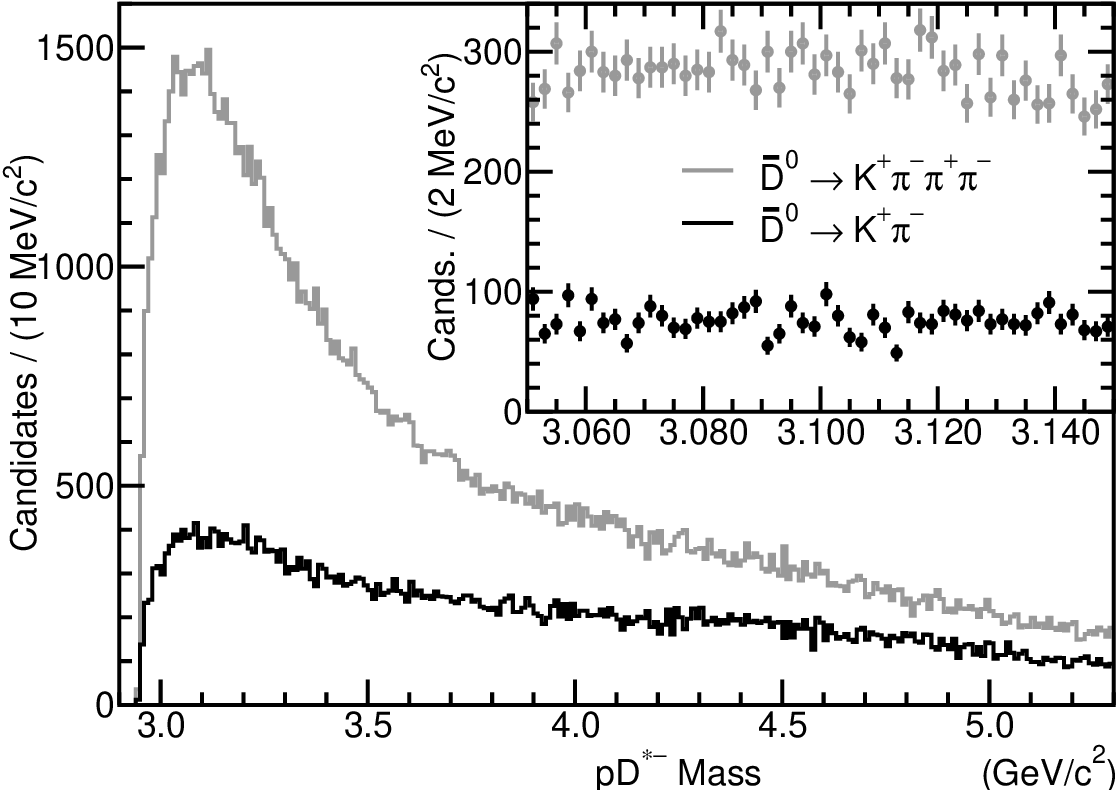}
\end{flushright}
\vspace{-0.7truecm}
\caption {
Invariant mass distributions for \Thc candidates in the data in
the (black) $K^+\pim$ and (gray) $K^+\pim\pim\pip$ decay modes, 
over a wide mass range and (inset) in the region near 3100 \mevcc.
 }
\label{fig:massall}
\end{figure}

We show \mpdst distributions for the \Thc candidates in the data
in Fig.~\ref{fig:massall} for the two \dbar0 decay modes.
They show no narrow structure; 
in particular they are smooth in the region near 3100 \mevcc, shown in
the inset, where the bin size is two-thirds of the resolution.
Corresponding distributions for sidebands in the \dbar0 and \dstm
masses and the mass differences show overall structure similar to
that in the signal region.
We consider several variations of the selection criteria that might
enhance a pentaquark signal, but in no case do we observe one.
To enhance our sensitivity to any production mechanism that
gives a \pstar spectrum different from that of the background,
we divide the data into nine \pstar ranges of width 500~\mevc covering 
values from 0 to 4.5~\gevc.
The background is lower at high \pstar, so we are more sensitive to 
mechanisms that produce harder spectra.
There is no evidence of a pentaquark signal in any \pstar range.

We quantify this null result by fitting a signal-plus-background
function to the \mpdst distribution in each \pstar range.
We use a $p$-wave Breit-Wigner lineshape convolved with the 
resolution function described above.
The RMS width of the reported \Thc signal is 12~\mevcc and consistent
with the H1 detector resolution~\cite{h1plb}.
Our mass resolution is considerably better, so we must consider a
range of possible natural widths $\Gamma$ of the \Thc.
We quote results for two assumed widths, $\Gamma = 1$~\mev, 
corresponding to a very narrow state, and $\Gamma = 28$~\mev,
corresponding to the width observed by H1, which we take as an upper limit.
For the background we use the function 
$f(m)\! =\! 0$ for $m\! <\! m_0$ and
$f(m)\! =\! \sqrt{1-(m_0/m)^2} \exp(a[1-(m_0/m)^2]) / m$ for $m\! >\! m_0$,
where $m_0\! =\! m_p+m_{\dstm}\! =$~2948~\mevcc is the threshold value and $a$
is a free parameter.
We fit over the range from threshold to 3300~\mevcc, 
except in the lowest \pstar range for the \ktpi mode.
Here the acceptance drops sharply near threshold 
and the fit range is restricted to the region above 3000~\mevcc.

\begin{figure}[t]
\begin{flushright}
\hspace*{-0.3cm}
\includegraphics[width=8.7cm]{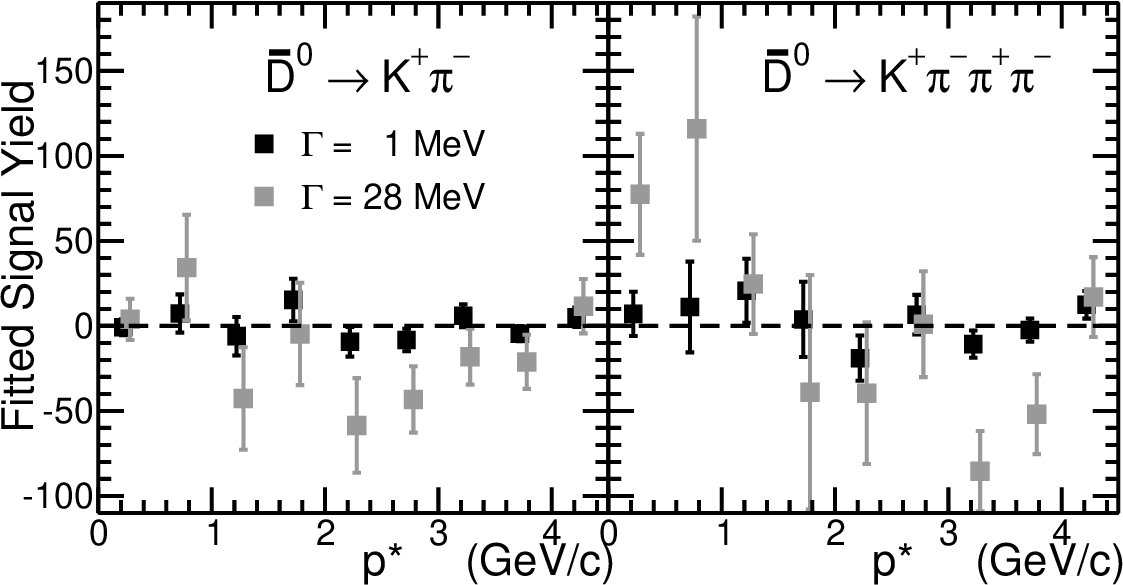}
\end{flushright}
\vspace{-0.7truecm}
\caption {\Thc yields from fits to the \mpdst
  distributions for the (left) $p\kpis$ and (right) $p\ktpis$ decay modes,
  assuming a mass of 3099~\mevcc and a natural
  width of $\Gamma\! =1$~\mev (black) or $\Gamma\! =$\hwid (gray). } 
\label{fig:yields}
\end{figure}

We perform maximum likelihood fits at several fixed \Thc mass values 
in the range 3087--3111~\mevcc.
In every case we find good fit quality and a signal amplitude 
consistent with zero.
We consider systematic effects in the fitting procedure by varying 
the signal and background functions and fit range; changes in the
signal yield are negligible compared with the statistical uncertainties.
The dependence on the assumed mass value is also small compared with the
statistical error in each case.
Fixing the mass to the reported value of 3099~\mevcc, we obtain the 
event yields shown in Fig.~\ref{fig:yields}.
There is no positive trend in the data, and the roughly symmetric scatter
of the points about zero indicates little
momentum-dependent bias in the background function.

\begin{figure}[t]
\begin{flushright}
\hspace*{-0.3cm}
\includegraphics[width=8.7cm]{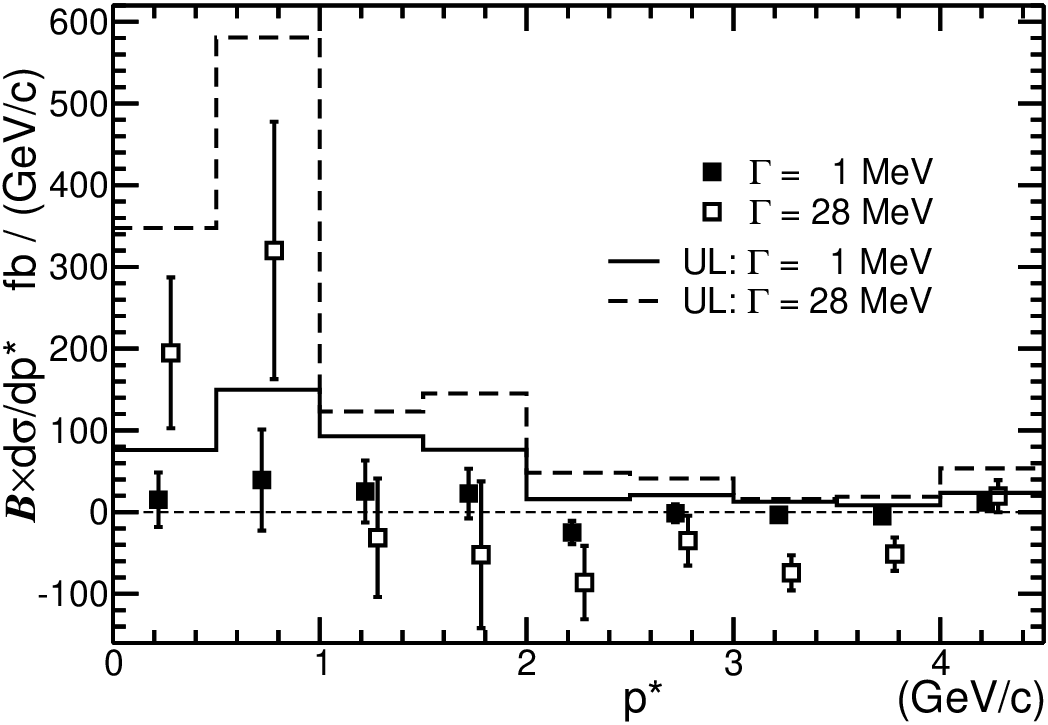}
\end{flushright}
\vspace{-0.4truecm}
\caption {
 Product of the \Thc differential production cross section and its 
 branching fraction to \pdstm (symbols) and corresponding $95\%$ CL
 upper limits (lines), 
 assuming natural widths of $\Gamma\! =1$~\mev (solid) and
 $\Gamma\! =\hwid$ (open/dashed), as functions of c.m.\ momentum.}
\label{fig:xsecul}
\end{figure}

In each \pstar range we divide the sum of the two signal yields 
by the sum of the two products of reconstruction efficiency and
$\dbar0 \!\!\to\!\kpi$ or $\dbar0 \!\!\to\!\ktpi$ branching fraction, 
the $\dstm\!\!\to\!\dbar0 \pis$ branching fraction,
the integrated luminosity, and the \pstar range.
This gives the product of the unknown $\Thc \!\!\to \!\pdstm$ branching
fraction, ${\cal B}$,
and the differential production cross section, \dsdpst.
The resulting values of \dsdotB for $\Gamma\! =$1~\mev and $\Gamma\! =$\hwid
are shown in Fig.~\ref{fig:xsecul}.
We derive an upper limit on the value in each
\pstar range under the assumption that it cannot be negative:
a Gaussian function centered at the measured value with RMS equal to
the total uncertainty is integrated from zero to infinity, and the
point at which the integral reaches 95\% of this total is taken as the limit.
These $95\%$ confidence level (CL) upper limits are
also shown in Fig.~\ref{fig:xsecul}.

We integrate \dsdotB over the full \pstar range from 0--4.5~\gevc,
taking into account the correlation in the systematic uncertainty,
to derive a total production cross section times branching 
fraction, \sdotB, for each of the two assumed $\Gamma$ values,
and calculate corresponding upper limits.
These limits are model independent; any postulated production spectrum
can be folded with the measured differential cross section to obtain a
smaller limit.
We calculate corresponding limits on the number of \Thc produced 
per $q\overline{q}$ ($q\! =\! udsc$) event and 
per $c\overline{c}$ event by dividing by the
respective cross sections for these types of events;
we also calculate a limit per \Y4S\ decay by integrating \dsdotB
over the range $\pstar\! <$~2~\gevc (the kinematic limit for $B$ meson
decays is 1.8~\gevc)
and dividing by our effective cross
section for $\epem \!\!\to\! \Y4S$.
These central values and limits are given in Table~\ref{tab:xsecul}.

\begin{table}[t]
\begin{center} 
\caption{
Total production cross section of the \Thc pentaquark candidate
times its branching fraction to \pdstm, \sdotB,
in $e^+e^-$ annihilations at $\sqrt{s}=$10.58 \gev,
for two assumed values of the natural width.
The corresponding 95\% CL upper limits on \sdotB and on ${\cal B}$
times the yields per \eeqq event, \eecc event, and \Y4S decay.
}
\vspace{0.2cm}
\begin{small} 
\begin{ruledtabular}

\begin{tabular}{lcc}
\vspace{0.1cm} 
                             & $\Gamma\! =1$~\mev & $\Gamma\! =\hwid$ \\ \hline
\sdotB   (fb)                & 40$\pm$44          & 102$\pm$111       \\
                             & $<$ 117            & $<$ 297           \\ \hline
${\cal B}\times$yield$\times 10^{-5}$ per &       &                   \\
\hspace*{0.3cm} \eeqq event  & $<$ 3.4            & $<$ 8.8           \\ 
\hspace*{0.3cm} \eecc event  & $<$ 8.5            & $<$ 22            \\ 
\hspace*{0.3cm} \Y4S decay   & $<$ 12             & $<$ 37            \\ 
\end{tabular} 

\end{ruledtabular}
\end{small} 
\label{tab:xsecul}
\end{center}
\end{table}

In summary, we perform a search 
in $e^+e^-$ annihilations at $\sqrt{s}=$10.58 \gev
for the pentaquark candidate state \Thc 
reported by the H1 collaboration.
We use the same decay mode as H1, $\Thc \!\!\to\! pD^{*-}$,
and find no evidence for the production of this state in a sample of
over 125,000 \pdstm combinations.
The components of this sample from $c$-quark fragmentation and 
$B^0/\Bbar^{0}\! +\! B^\pm$ decays are both at least 
100 times larger than the sample used by H1, 
implying that neither hard charm quarks nor $B$ mesons 
produced in deep inelastic scattering can be the source of the H1 signal.
We set upper limits on the product of the inclusive \Thc production cross 
section times branching fraction to this mode for two assumptions as
to its natural width, which are valid for any state in
the vicinity of 3100 \mevcc.
It would be interesting to compare these limits with the rate expected
for an ordinary charmed baryon of mass $\sim$3100~\mevcc.  
However rates have been measured for only two charmed baryons, the
$\Lambda_c^+(2285)$~\cite{RPP2004,bellelc} and
$\Sigma_c(2455)$~\cite{RPP2004}, with precision that does not allow a
meaningful estimate of the mass dependence.
The mass dependence observed~\cite{RPP2004} for non-charmed baryons
in \epem annihilations would predict a rate for a 3100~\mevcc baryon
about 1,000 times smaller than that of the $\Lambda_c^+(2285)$.
Our limits for a narrow state in both \eecc and \Y4S events are 
roughly 1,000 and 500 times below the measured $\Lambda_c^+(2285)$ and 
$\Sigma_c(2455)$ rates, respectively.
As a result the existence of an ordinary charmed baryon with this mass and 
decay mode cannot be excluded.

We are grateful for the excellent luminosity and machine conditions
provided by our \pep2\ colleagues, 
and for the substantial dedicated effort from
the computing organizations that support \babar.
The collaborating institutions wish to thank 
SLAC for its support and kind hospitality. 
This work is supported by
DOE
and NSF (USA),
NSERC (Canada),
IHEP (China),
CEA and
CNRS-IN2P3
(France),
BMBF and DFG
(Germany),
INFN (Italy),
FOM (The Netherlands),
NFR (Norway),
MIST (Russia), and
PPARC (United Kingdom). 
Individuals have received support from CONACyT (Mexico), 
Marie Curie EIF (European Union),
the A.~P.~Sloan Foundation, 
the Research Corporation,
and the Alexander von Humboldt Foundation.


\begin{thebibliography}{99}

\bibitem{thetarefs}
LEPS Collaboration, T.~Nakano \etal, 
Phys.\ Rev.\ Lett. {\bf 91},~012002~(2003). \\
SAPHIR Collaboration, J.~Barth \etal,
\newblock Phys.\ Lett.{} {\bf B~572},~127~(2003). \\
CLAS Collaboration, S.~Stepanyan \etal,
\newblock Phys.\ Rev.\ Lett.{} {\bf 91},~252001~(2003). \\
CLAS Collaboration, V.~Kubarovsky \etal,
\newblock Phys.\ Rev.\ Lett.{} {\bf 92},~032001~(2004).
\newblock Erratum; ibid, 049902. \\
DIANA Collaboration, V.V.~Barmin \etal,
\newblock Phys.\ Atom.\ Nucl.{} {\bf 66},~1715~(2003). \\
SVD Collaboration, A.~Aleev \etal, Yad.\ Fiz.\ {\bf 68}, 1012 (2005). \\
HERMES Collaboration, A.~Airapetian \etal,
\newblock Phys.\ Lett.{} {\bf B~585},~213~(2004). \\
A.E.~Asratyan, A.G.~Dolgolenko, and M.A.~Kubantsev,
\newblock Phys.\ Atom.\ Nucl.{} {\bf 67},~682~(2004). \\
COSY-TOF Collaboration, M.~Abdel-Bary \etal,
\newblock Phys.\ Lett.{} {\bf B~595},~127~(2004). \\
ZEUS Collaboration, S.~Chekanov \etal,
\newblock Phys.\ Lett.{} {\bf B~591},~7~(2004). 

\bibitem{prl92:042003}
NA49 Collaboration, C. Alt \etal,
\newblock Phys.\ Rev.\ Lett.{} {\bf 92},~042003~(2004).

\bibitem{h1plb}
H1 Collaboration, A. Aktas \etal,
\newblock Phys.\ Lett.\ {\bf B~588},~17~(2004).

\bibitem{review}
See, e.g., 
A.R.~Dzierba, C.A.~Meyer and A.P.~Szczepaniak, hep-ex/0412077 (2004) 
and references therein.

\bibitem{bbrlimits}
\babar\ Collaboration, B.\ Aubert {\em et al.},
Phys.\ Rev.\ Lett.{} {\bf 95},~042002~(2005). 

\bibitem{BABARdetector}
\babar\ Collaboration, B.\ Aubert {\em et al.},
Nucl.\ Inst.\ Meth.\ {\bf A~479}, 1 (2002).

\bibitem{jetset}
T.~Sjostrand, Comput.\ Phys.\ Commun.\  {\bf 82}, 74 (1994).

\bibitem{evtgen}
D.J.~Lange, Nucl.\ Inst.\ Meth.\  {\bf A~462}, 152 (2001).

\bibitem{GEANT}
GEANT4 Collaboration, S.~Agostinelli et al.,
Nucl.\ Inst.\ Meth.\  {\bf A~506}, 250 (2003).

\bibitem{bellelc}
Belle Collaboration, R. Seuster, et al., Phys.\ Rev.\ {\bf D 73},
032002 (2006).

\bibitem{RPP2004}
Particle Data Group, S.~Eidelman \etal,
\newblock Phys.\ Lett.~{\bf B~592}, 1  (2004).

\end{thebibliography}
\end{document}